\documentclass[a4paper]{jpconf}
\usepackage{graphicx}
\usepackage{epsfig}
\begin{document}
\def\b{\bar}
\def\d{\partial}
\def\D{\Delta}
\def\cD{{\cal D}}
\def\cK{{\cal K}}
\def\f{\varphi}
\def\g{\gamma}
\def\G{\Gamma}
\def\l{\lambda}
\def\L{\Lambda}
\def\M{{\Cal M}}
\def\m{\mu}
\def\n{\nu}
\def\p{\psi}
\def\q{\b q}
\def\r{\rho}
\def\t{\tau}
\def\x{\phi}
\def\X{\~\xi}
\def\~{\widetilde}
\def\h{\eta}
\def\bZ{\bar Z}
\def\cY{\bar Y}
\def\bY3{\bar Y_{,3}}
\def\Y3{Y_{,3}}
\def\z{\zeta}
\def\Z{{\b\zeta}}
\def\Y{{\bar Y}}
\def\cZ{{\bar Z}}
\def\`{\dot}
\def\be{\begin{equation}}
\def\ee{\end{equation}}
\def\bea{\begin{eqnarray}}
\def\eea{\end{eqnarray}}
\def\half{\frac{1}{2}}
\def\fn{\footnote}
\def\bh{black hole \ }
\def\cL{{\cal L}}
\def\cH{{\cal H}}
\def\cF{{\cal F}}
\def\cP{{\cal P}}
\def\cM{{\cal M}}
\def\ik{ik}
\def\mn{{\mu\nu}}
\def\a{\alpha}

\title{Stringlike structures in the real and complex Kerr-Schild geometry}

\author{Alexander Burinskii}

\address{Theor. Phys. Lab., NSI, Russian Academy of
Sciences, B. Tulskaya 52,  Moscow 115191, Russia}

\ead{bur@ibrae.ac.ru}

\begin{abstract}
Four-dimensional (4d) Kerr-Schild (KS) geometry displays remarkable relationships with
quantum world and theory of superstrings. In particular, the
Kerr-Newman (KN) solution has gyromagnetic ratio $g=2$, as that of the Dirac
electron and represents a consistent background for gravitational and electromagnetic
field of the electron.    As a consequence of the very big spin/mass ratio of spinning
particles, the black hole horizons disappear, obtaining the naked Kerr singular ring,
which displays close kinship with the Sen fundamental string solution to heterotic string
theory. The singular KN metric has to be regulated by a smooth rotating source, which sheds
the light on the consistent with gravity space-time structure of the dressed electron. We consider
the four-decade history of development of this structure which took finally the form of a
point-string-membrane-bubble complex which is reminiscent of the
enhancon model of  string/M-theory. We discuss also a complex string obtained
in the  complex structure of the Kerr geometry. It gives an extra dimension to the world-sheet
of the real Kerr string, forming a membrane by analogue with the  string/M-theory unification.
Finally, by analysis of the orientifold parity of the complex Kerr string, we obtain
that the determined by the Kerr theorem principal null congruence of the Kerr geometry  is
described by a quartic equation in the projective twistor space $CP^3 $, and therefore,
it creates the known Calabi-Yau twofold (K3 surface) in twistor space of the 4d KS geometry.
This strong parallelism of the 4d KS geometry with basic structures of string/M-theory
confirms important role of the KS geometry for quantum physics and allows us to assume
that complexification of 4d KS geometry may represent an alternative to  compactification
of higher dimensions. We connect this parallelism with mysterious N=2 superstring which has (complex) critical dimension two and may be embedded into complex KS geometry.  \end{abstract}

\section{Introduction}
It is now generally accepted that black holes are to be associated with elementary particles.
  New ideas and methods in the physics of black holes (BH) are based on the complex analyticity
and conformal field theory, that unites the physics
of black holes with superstring theory and particle physics. In spite of these evident
relationships, the path from superstring theory to particle physics represents an unsolved
problem, and moreover, it is constated that ``... realistic model of elementary particles still
appears to be a distant dream...'', John Schwarz \cite{Schw}.
Meanwhile, the four-dimensional Kerr-Newman solution for a charged
and rotating BH \cite{DKS} demonstrates in this relation some interesting surprises. Namely, it
 exhibits wonderful connections with spinning particles and simultaneously a remarkable
parallelism with basic structures of superstring theory.

In 1968 Carter noticed  that the KN solution has gyromagnetic ratio $g=2$, as that of the Dirac
electron \cite{Car}, and therefore, at least the asymptotic electromagnetic (em) and
gravitational fields of the electron should correspond to the KN solution with great precision.
Angular momentum of the KN solution $J=ma$ is proportional to mass $m$ and to Kerr's parameter
$a $, which is radius of the Kerr singular ring. The em and gravitational fields are
concentrated near the Kerr ring forming a type of waveguide, which causes its similarity
 with a closed string of the dual models. It has been assumed in \cite{IvBur,Bur0} that
 the em and gravitational traveling waves can propagate along the Kerr ring in analogue with
 excitations of the dual string models. Masses of  elementary particles are very small with
 respect to their spin, and the Kerr parameter of rotation $a =J/m ,$
 radius of the Kerr singular ring,
turns out to be of order of the Compton wave length of the electron $\sim \hbar/(2mc).$
In the dimensionless units $c=G=\hbar=1 $, the ratio $a/m$ is about $10^{44}$, which results in
disappearance of the BH horizons, and the  Kerr singular ring turns out to be naked.
Other spinning particles have approximately the same ratio $a/m$, and we arrive to trivial,
but very important fact that \emph{black holes cannot apparently be associated with spinning
particles}. The relevant to spinning particles geometry should be an over-rotating KN background
which  contains a closed string-like singularity.

It looks very bad, since instead of the consistent with quantum theory very  weak
gravitational field in vicinity of the electron, we reveal a singular core in the form of
a closed string, which represents a very strong defect of the background. One more trouble of
this result is a twosheeted structure of the resulting Kerr background, since
the naked Kerr singular ring represents a branch line of the Kerr space-time into two sheets.
This mysterious structure of the over-rotating KN geometry created the problems in
modelling of the source of the Kerr and KN solutions, which could consistently regulate this
background.
 Structure of such a source was specified step by step in many works
during more than four decades, in particular in the following papers
\cite{Keres,Isr,Bur0,IvBur,Ham,Lop,GG,BEHM,BurSol,BurQ}.  As a result, the
consistent regular source of the external KN solution, generating the
necessary very weak gravitational field in the vicinity of the source,
acquired the form of a relativistically rotating string-membrane-bubble complex:
a highly oblate bubble bounded by
an ellipsoidal membrane and by a circular fundamental string positioned on the
edge rim of the bubble.  The circular string appears as a result of regularization of
the Kerr singular ring, and we argue that traveling waves circulating along this string
should create a  singular point, D0-brane which exhibits ``zitterbewegung'' of the
Dirac electron. It was also shown \cite{BurQ} that wave excitations of the fundamental
Kerr string generate also
the axial half-strings which are going to infinity serving as carriers of de Broglie waves.

Along with circular fundamental string \cite{BurSol,IvBur,BurSen}, an open complex string was
obtained in the complex structure of the Kerr-Schild  geometry \cite{BurCStr1, BurCStr2}.\fn{Both
of these
strings  were rediscovered recently by Adamo and Newman  \cite{AdNew} by study of the asymptotic
form of the geodesic and shear-free congruences. They comment this discovery writing that ``...It would have
been a cruel god to have layed down such a pretty scheme
and not have it mean something deep.''} It should also be mentioned  that
the open Kerr string extends the world-sheet of the closed Kerr string to two-brane source,
displaying parallelism  with the process of string/M-theory unification \cite{BBS}, and moreover,
the resulting source of the KN solution is reminiscent in many details the enhancon model
 used by the superstring/M-theory unification \cite{LaJoh}. Finally, it has been
obtained recently \cite{BurAlter}, that the twistorial structure
of the four-dimensional KS geometry, being combined with the
orientifold structure of the complex Kerr string, creates a K3 surface (the
Calabi-Yau twofold) on the projective twistor space, which forms a holographic
skeleton of the KS space, determined by the Kerr theorem.
It confirms close relationships of the KS geometry with the basic structures of the superstring
theory and hints existence of some underlying theory which may
unify the Kerr's gravity with physics of elementary particles and
superstring theory. Twistor theory plays apparently principal role in
this unification, since the complex N=2 critical superstring
\cite{GSW} has also the inherent twistorial structure \cite{OogVaf,Gibb},
which allows it to be consistently embedded in the \emph{complex}
four-dimensional KS geometry. We start paper by treatment of the real, complex,
twistorial and
stringy structures of KS geometry along our previous papers
\cite{BurCStr1, BurCStr2, BurSol, BurAlter, BurNst}.

\section{Real structure of the Kerr-Newman (KN) geometry}

The KN metric is represented in the Kerr-Schild (KS) form \cite{DKS},
 \be g_\mn=\eta _\mn + 2h
e^3_\m e^3_\n \label{KSh} , \ee where $\eta_\mn$ is auxiliary
Minkowski background in Cartesian coordinates ${\rm x}= x^\m
=(t,x,y,z),$ \be h = P^2 \frac {mr-e^2/2} {r^2 + a^2 \cos^2
\theta}, \quad P=(1+Y\Y)/ \sqrt 2 ,  \label{h}\ee and $e^3 (\rm
x)$ is a tangent direction to a \emph{Principal Null Congruence
(PNC)}, which is determined by the form\fn{Here $ \z =
(x+iy)/\sqrt 2 ,\quad  \Z = (x-iy)/\sqrt 2 , \quad u = (z + t)/\sqrt 2
,\quad v = (z - t)/\sqrt 2, $ are the null Cartesian coordinates,
$r, \theta, \phi $ are the Kerr oblate spheroidal coordinates, and
$Y (\rm x) =e^{i\phi} \tan \frac{\theta}{2} $ is a projective
angular coordinate.  The used signature is $(-+++) $.} \be e^3_\m
dx^\m =du + \bar Y d \zeta + Y d \bar\zeta - Y\bar Y dv ,
\label{e3} \ee via function $Y (\rm x),$ which is controlled by
\emph{the Kerr theorem},
\cite{DKS,Pen,BurKerr1,BurKerr2}.

The twisting lightlike rays of the Kerr PNC are focussing in the
equatorial plane $\cos\theta=0 $, at the Kerr singular ring, $r=0$,
approaching it tangentially. As a result, the aligned with Kerr PNC metric and the KN
electromagnetic potential, \be A_{\m} = -P^{-2} {\rm Re} \frac {e} {
(r+ia \cos \theta)} e^3_\m \label{Amu} , \ee
 concentrate near the Kerr ring and form a closed lightlike gravitational waveguide
 \cite{IvBur}, playing the role of a closed string which may carry
 excitations in the form of the lightlike traveling waves \cite{Bur0,BurQ}, which
 take near the ring the form of pp-waves considered as string excitations in low energy string theory.
\begin{figure}[ht]
\centerline{\epsfig{figure=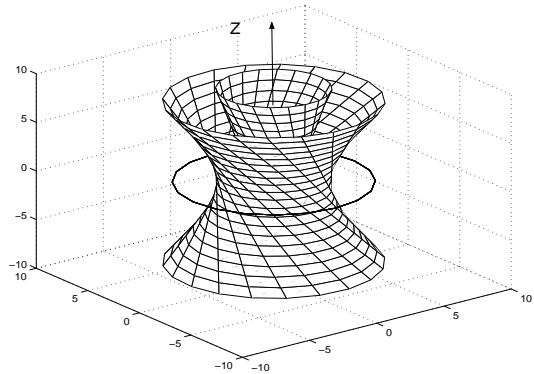,height=5cm,width=7cm}}
\caption{ Twistor null lines of the Kerr congruence are focused on
the Kerr singular ring which forms a branch line of space in two sheets.}
\end{figure}

\subsection{Regular source of the KN solution}
Gyromagnetic ratio of the KN solution is $g=2 $, as that of the
Dirac electron \cite{Car,DKS}. Therefore, the observable
parameters of the electron: mass, spin, charge and magnetic moment
indicate that the KN gravitational field corresponds to the
electron background geometry \cite{BurQ}, which created a series
of  works on the consistent with gravity classical model of the
electron \cite{Isr,Lop,Bur0,BurQ,DirKer,BurAxi,Beyond,TN}
and stimulated investigations of models of the source of the Kerr and
KN solution.

{\it (1) The Kerr ring as a closed string.} This model, created first in 1975 \cite{IvBur},
by analogue with the Dirac string of magnetic monopole, was supported then by the studies
of the fundamental string solutions to low-energy string theory \cite{DGHW,HorSteif}.
In 1992 Sen obtained two important
solutions to low-energy string theory: a) solution for fundamental
heterotic string \cite{SenFund}, and b) analogue of the Kerr
solution to low-energy string theory \cite{KerSen}.  It has
been shown then in \cite{BurSen} that  metric and electromagnetic
field near the Kerr ring in the solution (b) have the lightlike Killing structure similar
to that of the fundamental  heterotic string solution (a).
The only difference consisted in branching of the Kerr singularity related with two sheets of the  Kerr geometry.

{\it (2) String-membrane source.} Alternative line of investigations, related with
attempts to remove Kerr's twosheetedness was started by H. Keres \cite{Keres},
who noticed that the complex
radial distance $ \tilde r = r +i a \cos\theta, $ which determines
the KN gravitational and electromagnetic potential $ \Phi = {\rm Re} ( m
/\tilde r)$, being expressed in the Cartesian coordinates yields $
(\tilde r)^2 = x^2+y^2 +(z-ia)^2 $, and therefore, ``function
$\tilde r^{-1}$ and its first derivatives are continuous over all
space, beside the disk $z=0, \ x^2+y^2 \le a^2 , $'' and
``...  crossing the disk the derivatives have a jump discontinuity...''
indicating that the sources should lie on the disk. Similarly,
Israel truncated the Kerr  `negative' Kerr sheet along the disk $r=0 ,$
and showed in \cite{Isr} that  the resulting jump discontinuity
should create a matter distribution over the disk surface. He
interpreted the disk-like source of the Kerr geometry as a classical
model of electron. Next important step was done by Hamity \cite{Ham}, who
showed that Kerr's disk  has to be rigidly rotating, reaching the
velocity of light on the string-like border of the disk.
Finally, L\'opez \cite{Lop} constructed the regular source of the
KN solution, excising the singular region $r<r_e=e^2/2m$ (note that $r$ is the
Kerr ellipsoidal radial coordinate), and the Kerr-Newman (KN) regular source
took the form of a rigidly rotating ellipsoidal membrane, or the disk-like bubble,
flat interior of which is matched with the external KN solution along the bubble
surface $r=r_e ,$ being determined unambiguously by the equation
$ h(r) =0 $, where $h$ is given by (\ref{h}).

As a result, the gravitational singularity
disappears, and the space around the regularized disk-like source (= Compton region of
the dressed electron) turned out to be very close to flat. In the same time, the singular
electromagnetic field gets cutoff at $r=r_{reg}=r_e=e^2/2m $ (ellipsoidal  radial coordinate)
and is regularized. It turns the Kerr ring into regular fundamental string positioned
at the edge rim of the ellipsoidal bubble. As it was shown in \cite{BurQ}, the traveling waves
along circular string create the axial singular beams (outgoing pp-wave half-string)
which serve as carriers of de Broglie waves.

{\it (3) Gravitating soliton model.} In  \cite{BurSol}, the membrane-bubble model
was extended to a smooth field model based on the system of chiral superfields and on
the suggested in \cite{GG} modified form of the KN metric. The regular
 bubble-source was formed by a domain wall phase transition interpolating between
 the external KN background and a supersymmetric pseudovacuum state inside the bubble.
 The model represents a soliton of the oscillon type, since the internal false-vacuum
 state is built of the oscillating Higgs-like field. The KN vector potential is dragged by
 the relativistically rotating  string and forms a closed Wilson loop, which turns out
 to be quantized via interaction with the internal Higgs field, resulting in a consistent with
 KN gravity rotating soliton (for details see \cite{BurSol}).

  The  KN electromagnetic field is regularized and forms a  lightlike fundamental string
  positioned on the boundary of the disk-like bubble in equatorial plane $\theta=\pi/2$.
  New important effect is related with circular traveling waves which should propagate
  along fundamental string distorting the form of the bubble surface.
  The cutoff $r_e$ is determined by amplitude of the vector-potential, and there appear some nodes where
the cutoff $r_e$ approaches  zero, and regularization is to be broken.
The deformed by waves bubble boundary touches the Kerr singular ring creating singular points,
D0-branes which circulate along the bubble in equatorial plane.  These singular points
form the ends of the fundamental string and can be identified  with lightlike quarks or partons.
It turns out to be closely related with the fact that the fundamental circular string,
being parametrized by the light cone parameter $\sigma =t- n \phi $, cannot indeed be closed,
since the end points $\phi = 0$ and $\phi = \pi $ turn out to be disconnected by a time-like gap
$x^\m (t - 0) - x^\m (t - n \pi) ,$ which should be closed by the path of a massive  mode.
These end points are stuck at the bubble, which therefore acquires the role of a heavy quark
which closes this loop. It suggests a physical mechanism which can lie beyond
``zitterbewegung" of the Dirac electron. On the other hand, it resolves the seeming
contradiction between the point-like and extended electron: the naked electron is point-like
and represents a circulating singular pole (D0-brane), while the dressed KN electron exhibits
an unambiguously defined exhibits  an extended structure of the bag model, which is similar to the MIT and SLAC bag models, but is consistent with KN gravity.  From geometrical point of view, it represents a simplicial complex of the Compton size, consisting of the point, circular string,
ellipsoidal membrane and a false-vacuum bubble filled by Higgs field -- the complex system
of the D0-D1-D2-D3 branes.

{\it (4) Complex Kerr string.}
Along with the circular string structure, the Kerr-Schild geometry contains also a
\emph{complex string} \cite{BurCStr1, BurCStr2}, which appears in the
initiated by Newman complex representation of the Kerr geometry
\cite{LinNew}.  This string is open and parametrized by angular variable
$\theta \in [0, \pi]$. In fact, parameter $\theta$ extends the world-sheet of the
circular Kerr string (positioned at $\theta=\pi/2$) to the membrane source of the
L\'opez bubble model \cite{Lop,BurSol}. Structure of the regular bubble source of
the KN solution represents a copy of the enhancon model used for regularization
by the superstring/M-theory unification \cite{LaJoh}  and represents a counterpart
of the transfer from 10-dimensional superstrings to 11d $M$-theory \cite{BBS}.

\subsection{The Kerr theorem and twistorial structure of the KS geometry}
\textbf{The Kerr Theorem} \cite{DKS,BurNst,Pen} determines the shear-free null congruences
with tangent direction (\ref{e3}) by means of the solution $Y(\rm
x)$ of the equation \be F(T^A) =0 \label{F0KerrTeor}, \ee where
$F(T^A)$ is an arbitrary holomorphic function in the projective
twistor space   \be T^A= \{ Y, \quad \l ^1 = \z - Y v, \quad \l ^2
=u + Y \Z \} .\label{(TA)} \ee

Using the  Cartesian coordinates $x^\m ,$ one can rearrange
variables and reduce function $F(T^A)$ to the form $F(Y,x^\m),$
which allows one to get solution of the equation
(\ref{F0KerrTeor}) in the form $Y(x^\m).$

For the Kerr and KN solutions, the function $ F(Y,x^\m)$ turns out
to be quadratic in $Y,$ \be  F(Y,x^\m) = A(x^\m)  Y^2 + B(x^\m) Y +
C(x^\m), \label{FKN} \ee and the equation (\ref{F0KerrTeor})
represents
 a \emph{quadric }in the projective twistor space $ CP^3 $,
 with a non-degenerate determinant $\D = (B^2 - 4AC)^{1/2} $
which defines the complex radial distance \cite{BurKerr1,BurKerr2,BurNst}
\be \tilde r = - \D = -(B^2 - 4AC)^{1/2} . \label{trDet} \ee
 This case is explicitly resolved and yields two
   solutions \be Y^\pm (x^\m)= (- B \mp \tilde r )/2A, \label{Ypm}\ee
which allows one to restore two PNC by means of (\ref{e3}).
\noindent One can easily obtain from (\ref{FKN}) and (\ref{Ypm})
that the used in the metric (\ref{h}) and the em potential (\ref{Amu})
complex radial distance $\tilde r = r+ia\cos \theta$ may also be determined
from the Kerr generating function by the relation \be \tilde r = - dF/dY \label{tr} .\ee
Therefore, the Kerr singular ring, $\tilde r =0 ,$ is formed as a
caustic of the Kerr congruence, \be dF/dY=0 . \label{sing}\ee

As a consequence of  Vieta's formulas, the quadratic in $Y$
function (\ref{FKN}) may be expressed via the solutions
$Y^\pm(x^\m)$ in the form \be
F(Y,x^\m)=A(Y-Y^+(x^\m))(Y-Y^-(x^\m)) \label{FYYpm} .\ee
It is known, \cite{DKS}, that the linear in $Y$ generating functions $F$
correspond to the  light-like  solutions, while the subsequent analysis
of the generating functions of higher degrees in $Y$
showed that they split into products of the blocks of the first and second degrees in $Y ,$ and
therefore, they should correspond to multi-particle KS solutions \cite{Multiks1, Multiks2, Multiks3}.

\section{Complex Kerr geometry and an open complex string}

 One can see
 that the complex radial distance $\tilde r = r+ i a\cos\theta $ takes in
the Cartesian coordinates the form
\be \tilde r =\sqrt {x^2+y^2
+(z+ia)^2}, \ee
and therefore, the scalar component of the vector potential (\ref{Amu}) may be
obtained from the Coulomb solution $\phi(\vec x) = e/r = e/\sqrt{x^2+y^2
+z^2} $ by a complex shift $z\to z+ia ,$ or by  the shift of its singular
point $\vec x_0 =(0,0,0)$ in complex region $\vec x_0 \to (0,0, -ia).$
The complex shift was first considered  by Appel in 1887 \cite{App}, who noticed
that the Coulomb solution, being invariant solution to the linear Laplace equation
with respect to  real shifts of its origin $\vec x \to \vec x +\vec a ,$ should also be
invariant with respect to the complex shift. In spite of triviality of this procedure
from complex point of view, it yields
very nontrivial consequences in the real section, in particular, the singular point
of the Coulomb solution $\vec x_0 =(0,0,0)$ turns into singular ring
$x^2+y^2+(z+ia)^2 =0 $ (intersection of the sphere $x^2+y^2+z^2 =a^2 $ and the plane $z=0 $),
which becomes the branch line of the space into two sheets.

Newman showed that the KN electromagnetic field and the linearized KN gravitational field
 may be obtained from a complex retarded-time construction as the fields generated by a
 complex source propagating along a complex world line (CWL) \cite{LinNew}.
Later on, it has been shown that  linearized Newman's representation
acquires exact meaning in the Kerr-Schild class of metrics, \cite{Bur0, BurNst, BurKerr1, BurKerr2},
and the KN solution corresponds exactly to the field generated by a
{\it complex source propagating along  straight CWL} \be
x_L^\m(\t_L) = x_0^\m (0) +  u^\m \t_L + \frac{ia}{2} \{ k^\m_L -
k^\m_R \} ,\label{cwL}\ee where $u^\m=(1,0,0,0), \quad
k_R=(1,0,0,-1), \quad k_L=(1,0,0,1)$ and $\t_L=t_L+\sigma_L$ is a
complex retarded-time parameter. Index $L$ labels it as a
Left structure, and we should also add a complex conjugate Right
structure \be x_R^\m(\t_R) = x_0^\m (0) +  u^\m \t_R -
\frac{ia}{2} \{ k^\m_L - k^\m_R \} .\label{cwR}\ee Therefore, from
complex point of view the Kerr and Schwarzschild geometries are
equivalent and differ only by the relative position of their \emph{real slice}, which for
the Kerr solution goes aside of its center. Complex shift turns
the Schwarzschild radial directions $\vec n = \vec r /|r|$ into
twisted directions of the Kerr congruence, see Fig. 1.

Principal part of the complex retarded-time construction is a family of the
 complex light cones $\cK$ adjoined to CWL. Taking the Left CWL as a basis for treatment,
 one can represent the family of the adjoined complex light cones in spinor form
 $$ {\cK_L }= \{x: x =
x^{i}_L(\tau_L) + \psi ^{A}_{L} \sigma ^{i}_{A \dot { A}} \tilde{\psi
}^{\dot{A}}_{R} \} . \eqno(KL)  $$ \noindent The cones are split  into
two families of null planes: "left" $( \psi _{L}$ = const.; $\tilde{\psi
}_{R}$ -- var.) and "right"$( \tilde{\psi }_{R}$ = const.; $\psi _{L}$ -- var.).
These are the only two-dimensional planes which are wholly contained in the
complex null cone.
The rays of the principal null congruence of the Kerr geometry  are
the tracks of these complex null planes (right or left) on the real slice of
the Minkowski background. The real null direction $e^3 ,$ given by (\ref{e3}), is a
projective version of the spinor form  $ \psi ^{A}_{L} \sigma ^{i}_{A \dot { A}} \tilde{\psi
}^{\dot{A}}_{R}  ,$ expressed via the first projective twistor coordinate $Y=\psi_L^1/\psi_L^0.$
The real null direction $e^3$ is completed by two complex conjugate null forms
\be e^1 = d \zeta - Y dv, \qquad  e^2 = d \bar\zeta -  \bar Y dv.
\label{e12} \ee
One sees that the second projective twistor coordinate
$ \l ^1 =\z - Y v = x^\m e^1_\m $ represents
projection of the space-time point $x^\m$ on the null direction $e^1 ,$ while the third
 projective twistor coordinate $ \l ^2 =u
+ Y \Z = x^\m (e^3_\m- \Y e^1_\m) $ represents a linear combination of the projections on the
null directions $e^3$ and $e^1 .$ The determined by the Kerr theorem function $Y(x)$ allows one
to restore at each point $x$ the remaining twistor coordinates $ \l ^1$ and $ \l ^2$, and
to fix the incident Left null plane spanned by the null
directions $e^3$ and $e^1 .$ As a consequence, the
solution of the Kerr theorem $Y(x)$ foliates the Minkowski space-time into the Left null planes,
which due to specific structure of the KS metric (\ref{KSh}) turn out to be
null with respect to the curved KS space-time too, and therefore, the Kerr theorem performs a
specific linearization of the curved KS space-times, which justifies validity of the
twistor version of the Fourier transform \cite{Wit}.

Similarly, the  Right null planes
are to be spanned by the null directions $e^3$ and $e^2$. The twistor null rays of the Kerr
congruence are formed as intersections of the complex conjugate Left and Right null planes.

\subsection{Complex open string } It was obtained in
\cite{BurCStr1, BurCStr2, OogVaf} that the complex world line $x_0^\m (\t) ,$
parametrized by complex time $\t=t+i\sigma ,$ represents really a
two-dimensional surface which takes an intermediate position
between world-line of the particle and the string world-sheet.
The corresponding "hyperbolic string" equation \cite{OogVaf},
 $\d_\t \d_{\bar\t}
x_0(t,\sigma) =0 ,$ has the general solution \be x_0(t,\sigma)
= x_L(\t) + x_R(\bar\t) \ee as sum of the analytic and
anti-analytic modes $x_L(\t)$, $x_R(\bar\t)$, which are not
necessarily complex conjugate. The complex light cones $\cal K$ adjoined to each point
of the world-sheet $x_0^\m (\t) $ are split into the Left and Right null planes,
forming the linear N=2 structure of the complexified KS geometry.
The real KS structure appears as a result of \emph{projection of this structure on the real slice,
which is non-linear operation,} resulting in twisted structure of the real KS space-time.

\emph{For each real point} $x^\m ,$ the
parameters $\t$ and $\bar\t$ are to be determined by a complex
retarded-time construction. Complex source of the KN solution
corresponds to two \emph{straight} complex conjugate
world-lines (\ref{cwL}), (\ref{cwR}). Contrary to the real case,
the complex retarded-advanced times $\t^\mp = t \mp \tilde r $ may
be determined by two different (Left or Right) complex null
planes, which are generators of the complex light cone. It yields
four different roots for the Left and Right complex structures
\cite{BurKerr1, BurKerr2, BurNst} \bea \t_L^\mp &=&
t \mp (r_L + ia\cos\theta_L) \label{Lretadv} , \\
\t_R^\mp &=& t \mp (r_R + ia\cos\theta_R) \label{Rretadv}. \eea

The real slice condition determines the relation $\sigma = a \cos \theta$ , which connects
parameter $\sigma$ with angular directions of the Kerr congruence $\theta \in [0, \pi]
.$ It puts restriction $\sigma \in [-a, a] $ indicating that
\emph{the complex string is open}, and  its endpoints $\sigma =
\pm a$ may be associated  with the Chan-Paton charges of a
quark-antiquark pair. In the real slice, the complex endpoints of
the string are mapped to the north and south twistor null lines,
$\theta =0,\pi ,$ see Fig. 3.

\subsection{Orientifold projection} Boundary conditions of the complex open string
cannot consistently be set for its real and complex part simultaneously
\cite{BurCStr1, BurCStr2}. Solution of this obstacle requires an \emph{orientifold} structure \cite{BBS,GSW}
of the worldsheet of the open complex string. Orientifold
 turns the open string in a closed but
folded one. The world-sheet parity transformation $ \Omega: \sigma
\to - \sigma $ reverses orientation of the world sheet, and covers
it second time in mirror direction. Simultaneously, the Left and
Right modes are
 exchanged.\fn{Two oriented copies of the interval $\Sigma = [-a, a] ,$
$\Sigma^+ = [-a, a]$ and $ \Sigma^- = [-a, a]$, are joined,
forming a
 circle $ S^1 = \Sigma ^+\bigcup \Sigma ^-
$, parametrized by $\theta ,$ and map $\theta \to \sigma=a\cos
\theta $ covers the world-sheet twice.} The projection $\Omega$ is
combined with space reflection $R: r\to - r ,$ resulting in
$R\Omega: \tilde r \to -\tilde r ,$ which relates the retarded and
advanced folds \be R\Omega:  \t^+ \to \t^- \label{POmt} ,\ee
preserving analyticity of the world-sheet.
\begin{figure}[ht]
\centerline{\epsfig{figure=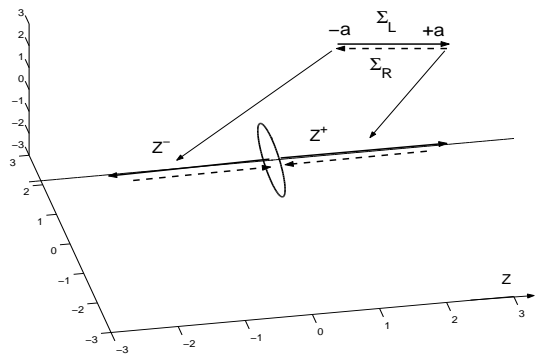,height=5cm,width=7cm}}
\caption{Ends of the open complex string, associated with quantum
numbers of quark-antiquark pair, are mapped onto the real
half-infinite $z^+, z^-$ axial strings. Dotted lines indicate
orientifold projection.}
\end{figure}
 The string modes $x_L(\t)$, $x_R(\bar\t)$ are extended on
the second half-cycle by the well-known extrapolation
\cite{BBS,GSW} \be x_L(\t^+) = x_R(\t^-); \quad x_R(\t^+) =
x_L(\t^-) , \label{orbi}\ee which forms the folded string, in which
the retarded and advanced modes are exchanged every half-cycle.

 The presented in Fig. 2 diagram shows a crossing symmetry of the four roots
$\t^\pm$ and $\bar\t^\pm ,$ for the complex retarded time, which
allows one  to replace the Right complex conjugate retarded-time
structure $x_R(\t^-)$
  by the antipodal Left advanced-time structure $x_L(\t^+),$ and works only in terms
  of the Left complex structures, omitting the index `L'. For more details on this antipodal relation see \cite{BurTMP1, BurTMP2}.

\section{Calabi-Yau twofold from the Kerr theorem}

\subsection{Problem of the non-stationarity}
The algebraically special exact KS solutions
of the Einstein-Maxwell equations obtained in the remarkable paper \cite{DKS}  have two
essential restrictions. First one is requirement of stationarity.
In particular, the Kerr and KN solutions  are stationary and their nonstationary
generalizations are unknown. In the KS formalism \cite{DKS} stationarity is characterized by a
constant Killing direction $K^\m$,
which corresponds to invariance of the metric $g_\mn$ with respect to action of the operator
$ \hat K = K^\m \d_\m ,$ i.e. $\hat K g^\mn =0 . $ Using Cartesian coordinates, one can express
$K^\m $  via coordinates of the straight CWL $x^\m_0(\t) ,$ complex source of the Kerr
and KN solutions, and obtain \be K^\m = \d_\t
x^\m_0(\t) .\label{Kmu}\ee
Stationarity of the metric implies stationarity of the Kerr congruence $\hat K e^3 =0 ,$ which
imposes the corresponding restriction on the coefficients $A,B,C$ of the generating function
of the Kerr theorem and the corresponding solutions \be \hat K Y =\hat K \Y =0 \label{KY} .\ee
All these functions turn out to be functions of the coordinates $x^\m_L(\t^-)$ and 4-velocity
of the CWL,
$u^\m(\t^-)=\dot x^\m_L(\t^-) \equiv \d_\t x^\m_L(\t)|_{\t^-} ,$ and finally they should
depend on the complex retarded time parameter $\t^- .$ \fn{Details of
  these  relations are not essential for our
  treatment here and can be found in \cite{BurNst, BurKerr1, BurKerr2}.}
  Although the determined by the Kerr theorem roots $Y^\pm(x,\t^-)$
  depend formally on the retarded  time $\t^- ,$ \emph{for the straight CWL} of the Kerr
  and KN solutions this dependence drops out from the final expressions, reducing
  to the dependence of the functions $F$ and $Y$ only on 4-velocity $u^\m(\t^-), $
  or on the related parameters $K^\m(\t)$ \cite{BurNst}.

The second restriction is related with idealization of the situation.

The non-stationary (accelerating) KS solutions are unknown. However, the known Kinnersley
solutions for accelerating non-rotating sources demonstrate that acceleration is always
accompanied by radiation. The obtained in \cite{BurA,BurPreQ} exact non-stationary em
solutions on the stationary KS background  show also that any em excitation of the Kerr
geometry creates specific radiation in the form of singular beams (pp-waves) propagating
along the null lines of the Kerr congruence. These beams are incoming on the negative sheet
of the Kerr geometry, and propagate towards the disk $r=0$,  when passing analytically through
the disk,  the beam-pulses become outgoing and propagate towards future infinity.
Some of the exact non-stationary em solutions (pp-waves) do not create additional
radiation and do not cause the recoil. However, these exact stationary solutions correspond
to an idealized situation with a congruence, which transfers analytically from the negative
to positive sheets of KS space-time and corresponds to an isolated stationary system without
internal and external sources.

When we consider the regularized KN solution, the idealized analytic KN background is to be
replaced by the background with a  membrane source separating the bulk of the
 positive Kerr sheet, $r \ge r_e ,$ from the bulk of the negative sheet, $r \le r_e .$
The membrane source  breaks analyticity of the Kerr solutions, and results in the scattering of
the pp-wave radiation  creating recoil and non-stationarity of the KS solutions.
In the moment of the collision of the beam pulse
with the membrane-source which closes the gate from negative to positive sheet  at $r=r_e ,$
there should appear a kink in the parameters of CWL.  Parameters of the Kerr ingoing
congruence $A_{in},B_{in},C_{in}$ should differ from parameters of
the outgoing one,
 $A_{out},B_{out},C_{out} .$ The in- and out-sources of the Kerr geometry
 turn out to be decorrelated and should be considered as independent sources, generating
 different congruences.
 In the simplest case, we can set $r_e=0,$
which corresponds to Keres-Israel-Hamity disklike membrane source
which blocks the gate from the negative to positive sheet, the
ingoing (real or virtual) photon will be scattered by the membrane and
produce a  recoil, causing a jump, $K^\m_{in} \ne K^\m_{out} ,$
described by a typical Feynman graph.
\begin{figure}[ht]
\centerline{\epsfig{figure=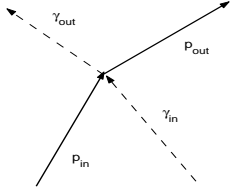,height=2.5cm,width=3cm}}
\caption{Recoil of the CWL caused by an electromagnetic beam or by
a photon propagating  from the past infinity through the Kerr ring
to the future infinity.}
\end{figure}

In the modern physics, treatment of the classical problem of
radiation reaction has reemerged in the connection with gravitational
radiation and problems of quantum gravity. In the seminal work by DeWitt and
Breme \cite{DeWitBrem}, the so-called `bi-tensor' fields and
Hadamard's `elementary solution' were considered as a useful
instrument for study of this problem in analogue  with the
two-point Green functions and the Feynman propagator.
One can note that orientifold represents a stringy analog of the
`bi-tensor' fields.

 Due to kink of the trajectory at the moment
$\t_0 ,$ the relationship between the retarded and advanced roots
is to be be biased (as it is shown in the diagram of Fig. 2), and the
 retarded coefficients of the Kerr generating
 function $A(\t^-), \ B(\t^-), \ C(\t^-)$ should not match in general with
 the coefficients $A(\t^+), \ B(\t^+), \ C(\t^+)$,
corresponding to advanced time $\t^+ .$ As a result,  the retarded and
advanced world lines, $x^\m_L(\t^-)$ and $x^\m_L(\t^+),$ generating two
different functions, $F^+$ and $F^- ,$ and correspondingly, two different
twistorial manifolds have to be determined by a two-particle version
of the Kerr theorem.

In accordance with the treatment of the multi-particle Kerr-Schild solutions
\cite{Multiks1, Multiks2, Multiks3},  generating function of the two-particle system has to be
 determined  by the function  $ F^{(2)}(T^A) = F^+(T^A) F^-(T^A)
$, composed as a product of two functions  $F^+$ and $F^-$
corresponding to the retarded, $x_L(\t^-)$, and advanced,
$x_L(\t^+)$, complex sources. The function \be F^{(2)}(T^A, x_L^+,
x_L^-)=F^+ (T^A, x_L^+)F^- (T^A, x_L^-) \label{F2} \ee represents
a bi-scalar, or two-point function associated with  the
world-sheet parity, $\t_L^- \leftrightarrow \t_L^+ .$

Both factors, $F^+$ and $F^-$ are quadratic in $T^A ,$ and
each of the partial equations $F^+=0$ (or $F^-=0$) generates a
\emph{quadric} in the projective twistor space $CP^3 $ corresponding to
the usual two-sheeted structure of the Kerr geometry with two different real Killing
directions consistent with the Debney-Kerr-Schild formalism.

The `product' manifold, determined by the equation $F^{(2)}(T^A,
x_L^+, x_L^-)=0 ,$ corresponds to a fourfold described as \emph{a
quartic} in the projective twistor space $ CP^3 $, which is the
Calabi-Yau (complex) twofold, or the  well-known K3 surface
used in diverse models of the string compactification and also by
 generalization of superstring theory to
M-theory \cite{BBS,GSW}. We obtain here that
dynamical generalization of the Kerr
geometry requires splitting of the complex source of the Kerr
geometry into independent retarded and advanced components described
by the orientifold parity of the world-sheet, and
application of the Kerr theorem creates the K3 surface in the
projective twistor space $CP^3 $.

\subsection{Matching of the retarded and advanced fields}
One sees that the fields at a fixed \emph{ real} point $x^\m$ near
the KN source may be represented as a sum of the retarded field
generated by the past complex light cone (root $\t^-$) and the
advanced field generated by the future complex light cone (root
$\t^+$).  For the stationary case, these points lie on the same
straight CWL and generate two congruences of the same 3d form,
which differ by their space-orientation. The  generated by $\t^-$
congruence is \emph{outgoing} on the physical (positive) sheet,
$r>0$, while the congruence generated by  $\t^+$ is formed by the
\emph{ingoing} rays, and therefore, its physical sheet is
`negative', and corresponds to $r<0 $ for the root $\t^+ .$ So far as the momenta
of the complex Kerr source $u^\m_{in}$ and $u^\m_{out}$ are to be
real,  there may be set the time-ordering $ \t^- \prec \t^+ ,$
which shows that the retarded physical sheet precedes the sheet of
advanced fields.

The four sheets of these congruences may be marked by the four
solutions $Y^\pm(x,\t^+)$ and $Y^\pm(x,\t^-).$ For the stationary
KN solution, the Kerr generating function $F(Y, x(\t))$ is
independent of $\t ,$ and as a result $F^+(Y) = F^-(Y),$ and the
four congruences coincide pairwise,
 $Y^\pm(x,\t^+) = Y^\mp(x,\t^-),$ forming
 a \emph{global} orientifold parity for the stationary Kerr geometry.

 The orientifold condition (\ref{orbi})  matches the lines of Kerr
 congruences at the world-sheet of the complex Kerr string.
 On the real slice of the Kerr geometry, it induces matching of the
 Kerr congruence at the membrane-surface of the regularized KN solution
 which separates the external KN solution from the internal false-vacuum state.
As we discussed in Sec. 2, in the simplest Keres-Israel-Hamity model
of the KN source, this membrane represents the disk $r=0 ,$ which separates
the `positive' and `negative' sheets of
the Kerr geometry. The closed lightlike Kerr string lies at the boundary of the disk at
$r=0$ and $\cos \theta = 0 ,$ and this string is extended to membrane by the extra
world-sheet parameter $\theta \in [0, \pi ] $. One sees
here close parallelism with the superstring/M-theory
correspondence, where the heterotic string transfers to a membrane
growing by the eleventh dimension \cite{BBS}.

 The source of the  \emph{gravitating soliton} model represents a false-vacuum bubble,
  boundary of which represents a domain wall membrane providing smooth  phase transition from external KN solution to a flat regular internal space. The membrane-boundary of the bubble has ellipsoidal form determined by the surface $r=r_{reg},$ and therefore, it fixes the
  boundary of analyticity of the KN geometry, or the boundaries of the `physical sheets' for the advanced and retarded fields
 \be \t^-_L|_{r=r_{reg}} = t-r_{reg} -ia\cos\theta \equiv t -(r_{reg}+ia\cos\theta)
 = (\t^-_R)|_{r=r_{reg}}^\ast = \t^+_L|_{r=r_{reg}}. \label{matchorbi} \ee
It means that analyticity of the Kerr congruence is to be broken on the
boundary $r=r_{reg} ,$ where the retarded `outgoing' Kerr congruence should be replaced
by the `ingoing' advanced congruence. As a result the both retarded and advanced fields
should be taken into account for the domain wall boundary conditions at $r=r_{reg} .$

However, the  antipodal relation $Y^+ =- 1/\bar Y^- $  between
 the \emph{retarded} KN solution on the positive physical sheet $Y^+(x,\t^-)$ and
 the \emph{advanced} conjugate solution $\bar Y^- $ on the negative sheet (see \cite{BurTMP1, BurTMP2}) allows us to consider these solutions as analytically related, displaying close analogue with formation of the Feynman propagator from a unique analytic multisheeted  solution.

Using this analogue, one can define the physical space-time of a spinning particle by
analytical connection of two of the four physical sheets of the K3 surface.
For example,  unification of the positive
sheet of the retarded $F^-$-solution $Y^+(x,\t^-)$ with negative sheet of the advanced source
(corresponding to $F^+ $) may be considered as a physical space-time of some spinning particle,
while the  `physical' spacetime for the corresponding
\emph{anti-particle} could be identified with alternative choice of two physical sheets:
positive sheet solution $Y^-(x,\t^+)$ of the advanced generating function $F^+ ,$ extended
analytically on the negative sheet of the retarded generator $F^-$.

\section{Outlook}
One sees that the Kerr-Schild geometry displays
striking parallelism with basic structures of  superstring theory.
In particular, one of the striking results is the presence of
inherent Calabi-Yau twofold in the complex twistorial structure of
the Kerr geometry. In the recent paper \cite{BurQ} we argued that
it is not accidental, because gravity is a fundamental part of the
superstring theory, and the  Kerr-Schild gravity, being based
on the twistor theory, Kerr theorem and the wonderful multisheeted
complex Kerr geometry \cite{Multiks1, Multiks2, Multiks3}, displays the most
deep inherent  structures of space-time which may lie beyond quantum physics.

In many respects the Kerr-Schild gravity represents a version of
Witten's twistor-string theory \cite{Wit}, which is also
four-dimensional, based on twistors and related with experimental
particle physics. It has been observed recently \cite{BurTMP1,BurTMP2}
that the complex Kerr string has much in common  with the N=2 critical superstring
\cite{OogVaf,GSW,DAddaLiz}, which is also related with twistors. The N=2 string is
complex and has the complex critical dimension two, which corresponds to four
real dimensions,  indicating that it may lead to
four-dimensional theory of superstrings. However, signature of the N=2 string may only be
(2,2) or (4,0), which caused  obstacles for embedding of this
string in the space-times with Minkowskian signature. Up to our
knowledge, the trouble was not resolved, and the initially
enormous interest to N=2 string seems to be dampened. Meanwhile,
embedding of the N=2 string in the complexified Kerr geometry is
very natural and hints that stringlike structures of the
real and complex Kerr geometry may be related with
 underlying dynamics of the N=2 superstring \cite{BurAlter,BurTMP1}.

Along with wonderful parallelism with the standard superstring
theory, the stringy system of the four-dimensional KN geometry
displays very essential peculiarities. First of all note
that the discussed regular source of the KN solution provides consistent
matching with quantum theory, since it leads to a flat space-time
in the Compton region and predicts the Compton size of the dressed electron
and some other of its remarkable features  \cite{BurSol}.
Among them is the discussed above closed Kerr string positioned on the
boundary of the Compton region. This string is lightlike and similar to
Discrete Light Cone Quantization (DLCQ) circle of M(atrix) theory \cite{Suss}.
It has been noted in \cite{BurOrbi}
that such a string cannot be closed indeed. This circular string should have the
end points (D0-branes) which may be in the same space position, but have to be
separated by a time-interval. It is supported by the discussed in Sec. 2 singular pole
which appears as a consequence of the wave excitations of the gravitating soliton model.
Therefore, along with the stringy D1-brane of the KN source, there appears the adjoined
lightlike D0-brane (which is indeed a pair of D0-brane-antibrane system), and the
extended source takes the form of a complex of the D0-D1-D2-D3 branes.
The singular lightlike D0-component describes zitterbewegung of the
point-like electron, while the complex  of D1-D2-D3-branes determines its extended
dressed shape \cite{BurQ}.

The considered stringy structures of the real and complex Kerr
geometry set a parallelism between the 4d Kerr geometry and
superstring theory, indicating the potential role of this alliance
in the particle physics. On the other hand, these relationships
show that complexification of the Kerr geometry may serve as an
alternative to traditional compactification of higher dimensions
in superstring theory.

{\bf Acknowledgements.} Author thanks T.~Nieuwenhuizen,
D.~Gal'tsov, K.~Stepaniants and L.~J\"arv for useful discussions and
comments, and also P.~Kuusk for kind invitation to this
conference and financial support. This work was supported by RFBR
grant 13-01-00602.

\end{document}